\documentclass[12pt,linesnumbered,ruled,vlined]{article}
\usepackage{mathptmx}
\usepackage[utf8x]{inputenc}
\usepackage{color}
\usepackage[table,xcdraw]{xcolor}
\usepackage{tabularx}
\usepackage{natbib}
\usepackage{array}
\usepackage{booktabs}
\usepackage{mathtools}
\usepackage{amsmath}
\usepackage{amsthm}
\usepackage{amssymb}
\usepackage{graphicx}
\usepackage{multirow}
\usepackage{lscape} 
\usepackage{makecell} 
\usepackage{threeparttable}
\usepackage[unicode=true,pdfusetitle,
 bookmarks=true,bookmarksnumbered=false,bookmarksopen=false,
 breaklinks=false,pdfborder={0 0 1},backref=false,colorlinks=true]
 {hyperref}
\hypersetup{
 pdfborderstyle={},pdfborderstyle={},pdfborderstyle={},urlcolor=blue,linkcolor=blue,citecolor=blue}

\makeatletter

\@ifundefined{date}{}{\date{}}

\usepackage{amsfonts}
\usepackage{dsfont}

\numberwithin{equation}{section}

\allowdisplaybreaks[4]
\DeclareMathAlphabet{\mathcal}{OMS}{cmsy}{m}{n}

\makeatother

\usepackage{authblk}
\usepackage{tikz}

\begin{document}


\title{Sino-US S\&T Frictions and Transnational Knowledge Flows:
Evidence from machine learning and cross-national patent data}
\author[1,2]{Yanqing Yang\thanks{yanqingyang@fudan.edu.cn}}
\author[2]{Nan Zhang\thanks{ 22110680026@m.fudan.edu.cn}}
\author[3]{Jinfeng Ge\thanks{jfge@shmtu.edu.cn}}
\author[1]{Yan Xu\thanks{xu.yan.fudan@outlook.com}}
\affil[1]{Shanghai Academy of AI for Science}
\affil[2]{Fudan University}
\affil[3]{Shanghai Maritime University}

\date{\today}

\maketitle


\begin{abstract}
\vspace{-0.1cm}
This paper identifies the impact of China-US science and technology (S\&T) friction on knowledge flows in different  fields, using data on invention patent applications from China, the US, Europe, and the World Patent Office (WPO) along with machine-learning-based econometric methods. The empirical results find that the negative impacts of China-US S\&T frictions on cross-border knowledge flows are confined to a limited number of technology areas during the period of our observation. This paper further explores the characteristics of the negatively impacted technology areas, and the empirical results show that technology areas that rely more on basic scientific research, where the distribution of the US scientific and technological strength is more concentrated, and where the gap between U.S. and Chinese science and technology is narrower, are more likely to be the victims of the Sino-US S\&T friction.

\end{abstract}
	\vspace{-0.3cm}
\smallskip
	
 \smallskip


\section{Introduction}
The core insight proposed by endogenous economic growth theory (Romer, 1990) is that technological progress is the most important source of long-term economic growth. As demonstrated by patents, today, China and the United States are the two most important economies driving technological progress globally. According to statistics from the World Intellectual Property Organization (WIPO), in 2021, the combined valid patents of China and the United States accounted for 42\% of the total global patents.\footnote{Among them, China accounts for 22\% of the global total, while the United States accounts for 20\%.} The technological flow between China and the United States plays a crucial role in global technological diffusion and knowledge flow. From 2015 to 2020, during a six-year period, cross-border patent applications between China and the US reached 400,000, equivalent to half of the total valid patents in Germany’s history.

Although the knowledge flow between China and the United States has promoted economic development in both countries and globally, China’s rapid technological catch-up has triggered growing concerns in the United States about its global technological leadership. In August 2017, the Trump administration initiated the ``301 Investigation" against China, and the US began implementing large-scale restrictive policies towards China’s technology sector. To constrain China’s technological development, the US government introduced a series of substantive restrictions on technology transfer and market access, spanning tariffs, investment, and the movement of technical personnel. Within the policy discussions in the US regarding technological and economic relations with China, especially on technology restriction strategies, there are two strands of viewpoints. One is the so-called comprehensive ``decoupling" policy, which seeks to completely sever technological ties between China and the US, halting the supply of technology to China while preventing Chinese tech companies from competing in the US market.\footnote{ In the technological field, the decoupling measures against China during the Trump administration mainly included the following aspects: First, three lists: the Entity List and the Military End User List, which were used to restrict the export of technology products to China; and the Communist Chinese Military Companies List, which was used to prohibit stock trading with Chinese companies. Second, two executive orders: the first one on August 6, 2020, prohibited transactions in the US with WeChat, TikTok, and their parent companies, and also banned updates and downloads of these two apps on smartphones; the second one on August 14, 2020, required ByteDance to divest TikTok based on the recommendation of the Committee on Foreign Investment in the United States (CFIUS). Both of these executive orders were not implemented due to court rulings and expiration of deadlines. Finally, there were a series of measures regarding 5G and communications, including launching the ``Clean Network" initiative in three key areas: operators, cloud services, and submarine cables; studying the revocation and termination of China Telecom’s operating license in the US; removing "national security-threatening" communications equipment and services from rural areas; and lobbying other countries to abandon Chinese-made 5G equipment.} The other is the so-called ``small yard, high fence" policy, originally proposed by the US think tank researcher Samm Sack \citep{sacks2019testimony}. It targets specific technology fields, aiming to block and constrain China’s potential technological threats in certain ``small yard" areas, building ``high fences" around them through various policies. On April 27, 2023, US National Security Advisor Jake Sullivan delivered a speech at the Brookings Institution, clearly proposing the ``small yard, high fence" strategy to protect US critical technologies, including restricting the export of the most advanced semiconductor technologies to China.

Regarding the economic impact of the Sino-US trade friction that began with the 2017 ``301 Investigation", extensive research has been conducted, offering different insights into the impacts of Sino-US trade friction on global value chains, supply chains, and corporate innovation. However, there is limited literature on the specific effects of US trade tariffs and technological restrictions on the flow of technology between China and the US. This paper studies Sino-US technological friction from the perspective of technology diffusion and policy impact heterogeneity, aiming to gain a deeper understanding of its effects on technology and knowledge flow between China and the US.

This paper uses patent application data from patent offices of multiple countries and regions worldwide, and employs machine learning methods to identify the technology fields impacted by policies among the nearly 40,000 technology fields defined by international patent classification numbers (IPC). From a statistical perspective, this paper adopts the C-Lasso machine learning model proposed by \cite{su2016identifying}, using event study methodology to identify the heterogeneous policy effects on different technological fields. 

Empirical results from this study show that, although the overall impact of Sino-US technological friction on technology knowledge flow between China and the US is not significant, after identifying the heterogeneous impacts of Sino-US technological friction on different technological fields, we find that the effects of policy shocks are limited to certain technology fields, and the negative impacts are highly significant. In short, the empirical results support that the nature of Sino-US technological friction is indeed ``small yard, high fence", rather than a comprehensive ``decoupling". We further examine the knowledge flow between China and the US in terms of technology inflow and outflow, and the empirical results show that the overlap of technology fields affected by negative shocks in these two dimensions is low, indicating that the policy shocks in these two dimensions stem from different policy combinations, further highlighting the multidimensional and complex nature of Sino-US technological friction.

Using the US invention patent database on reliance on basic science (Reliance on Science) compiled by \cite{marx2020reliance}, and citation network data of patents from seven major technological nations, this paper further explores the determining factors of the technology fields affected by policy shocks. The empirical results show that technology fields more reliant on basic scientific research, with stronger US technological strength and smaller technological gaps between China and the US, are more likely to be negatively impacted in Sino-US technological friction.

The contributions of this paper are as follows: First, based on global patent data, it uses data-driven AI methods to measure the heterogeneity of technology field knowledge flow between China and the US under the influence of policies, clarifying the essential characteristics of US technology policies towards China. Second, through the C-Lasso machine learning method, it identifies the technology fields affected by heterogeneous policy shocks with high granularity, not only qualitatively confirming the technology fields impacted by Sino-US technological friction, but also quantitatively analyzing the impact of these policies on technology knowledge flow between the two countries, filling a gap in existing research. Finally, the study on the factors affecting technology fields impacted by policy shocks reveals the basis for the US’s strategy of technological constraints on China, providing important policy references for China in responding to US technological restriction policies.

\section{Literature Review and Research Background}
\subsection{Literature Review}
This paper is primarily related to three strands of economic literature: the first is the literature on geopolitical impact on the economy; the second is the literature discussing cross-border knowledge flow and innovation; and the third is the literature combining machine learning methods with econometric techniques.

Since the outbreak of the China-US trade friction in 2018, literature on the economic effects of geopolitical factors on various countries’ economies has emerged rapidly. \cite{aiyar2023geo}and other working papers from the International Monetary Fund (IMF) have seriously discussed the economic fragmentation caused by geopolitical factors and expressed concerns about the future multilateral global system. Existing literature has studied the impact and effects of China-US trade friction (\citealp{li2018effectiveness,wei2019sino,li2022sino,cui2022measuring}), analyzing the impact of the China-US trade friction on the Chinese economy. \cite{amiti2019impact} and \cite{fajgelbaum2022economic} have reviewed literature on the impact of the trade friction on China-US economic relations, particularly the US economy. Of course, the literature most relevant to this paper focuses on the impact of China-US technological friction on the economies of both countries. \cite{han2021mapping} studied the integration and exchanges between China and the US in the technology field over the past two decades, as well as the changes in the degree of technological interdependence, measuring the role of technological integration in promoting China’s technological progress. Based on these findings, the study further points out that technological decoupling between China and the US will significantly hinder their technological integration and exchanges, severely obstructing technological development in both countries.

The global dissemination of knowledge has advanced the global technological frontier, narrowing the technological gap between countries. In an open economy, technological diffusion and knowledge flow are crucial elements in determining global economic development \citep{berkes2022global}. International openness promotes domestic competition \citep{buera2020global}, boosts international trade (\citealp{melitz2021trade,akcigit2022international}), and thus improves knowledge flow and technology adoption between countries. A series of Chinese-language literature has explored how knowledge inflows or technology diffusion from developed countries to China affect domestic technological progress or productivity (\citealp{jiang2019impact,qu2019technological,xiao2016foreign}). This paper, however, studies the impact of geopolitical friction on knowledge flow between China and the US from the perspective of China-US technological friction.

Machine learning methods provide a standard approach for non-parametric searches for heterogeneity and are an attractive option for estimating heterogeneous causal inference \citep{athey2017state} . These methods also provide tools to minimize overfitting, thereby maximizing out-of-sample predictive power. Traditionally, machine learning algorithms were designed for prediction \citep{mullainathan2017machine} , rather than for causal inference and parameter estimation. In the last decade, research using machine learning algorithms for causal inference has been very active \citep{chernozhukov2018generic} . The method adopted by \cite{su2016identifying} in this paper is a typical approach combining machine learning and econometrics.

\subsection{Research Background}
In August 2017, the US Trade Office launched an investigation into China’s technology transfer, intellectual property, and innovation-related practices and policies under Section 301 of the Trade Act of 1974. The US unilaterally claimed that China had damaged US intellectual property interests in China through market access restrictions and technology transfer requirements. In January 2018, the US government formally began imposing tariffs on Chinese goods, thus initiating the China-US trade conflict.\footnote{On January 22, 2018, the US government announced global safeguard measures for imports of large washing machines and photovoltaic products, imposing tariffs of 30\% and 50\% for periods of 4 years and 3 years, respectively.} 

During the previous Trump administration, there are three aspects of the decoupling measures in the field of technology. First, the three lists: the Entity List, the Military End User List, and the Communist Chinese Military Companies List. The first two lists were used to restrict the export of technology products to China, while the third was used to prohibit stock trading with Chinese companies. Second, two executive orders: the first one on August 6, 2020, prohibited transactions within the US involving WeChat, TikTok, and their parent companies, and banned updates and downloads of these two apps on smartphones; the second one on August 14, 2020, required ByteDance to divest TikTok based on a recommendation from the Committee on Foreign Investment in the United States (CFIUS). Both executive orders were not implemented due to court rulings and the expiration of deadlines. Finally, there were a series of measures regarding 5G and communications, including the ``Clean Network" initiative across three major areas: operators, cloud services, and submarine cables; examining whether to revoke and terminate China Telecom’s operating license in the US; removing ``national security-threatening" communications equipment and services from rural areas; and lobbying other countries to abandon Chinese-made 5G equipment.

Key events affecting China-US tech companies include: in May 2019, President Trump issued a national security order placing Huawei on the Entity List. Regarding export control, in 2018, the US enacted the Export Control Reform Act, which specifically restricted the export of ``emerging and foundational technologies" (the Biden administration has gradually focused on strengthening multilateral export controls against China, amending the Wassenaar Arrangement, and mobilizing allied countries to participate in blocking China’s high-tech development). In terms of investment restrictions, in 2018, Trump signed the Foreign Investment Risk Review Modernization Act, allowing the US Foreign Investment Committee to review Chinese state-backed investments in US emerging tech companies and veto such investments under the guise of ``national security", especially Chinese investments in high-tech fields. 

In October 2018, Samm Sacks, a senior researcher at the New America think tank, first proposed the ``small yard, high fence" strategy for China’s technological defense. A group of American scholars, led by Samm Sacks, argued that building high fences around a ``small yard" would help regulators more effectively screen activities within the ``small yard" while minimizing collateral damage to adjacent high-tech fields. The ``small yard, high fence"  strategy gradually gained favor in the US Congress as a favored approach to China’s technology policy. On April 27, 2023, Jake Sullivan, the US National Security Advisor, explicitly confirmed the ``small yard, high fence" strategy in a speech at the Brookings Institution. Sullivan argued that these were carefully tailored measures, but contrary to China’s view, this was not a ``tech blockade" targeting emerging economies, but rather specific technologies and countries intending to challenge the US militarily. 

The policy logic of the ``small yard, high fence" is clear: the US needs to identify specific technologies and research areas directly related to US national security (the ``small yard") and set appropriate strategic boundaries (the ``high fence"). Core technologies within the ``small yard" should face stricter technology restrictions against China, while other high-tech fields outside the ``small yard" could be reopened for China. However, the precise scope of the ``small yard" has not yet been publicly outlined by the US. In 2019, Samm Sacks testified before the US Senate and proposed that high-tech and emerging technologies meeting three key criteria should be included in the ``small yard" restrictions, which were later endorsed by the China-US Technology Relations Working Group as a starting point for further discussions. Specifically, the three criteria are: the technologies are crucial for military purposes; China has relatively little knowledge of the technology; and the US is at the forefront of the technology’s development. However, in practice, the US has not announced a clear policy regarding the selection of the ``small yard", indicating the ambiguity of its technology policy toward China.

Furthermore, the policy sources behind China-US technological friction are diverse, complex, and ambiguous. First, many policies affect the technological sector. \cite{luo2020mapping} in their report on Mapping US-China Technology Decoupling explored the impact of policies on China-US technology from various dimensions such as import and export controls, cross-border data flow restrictions, supply chain reviews, multinational investments, travel and work permits, etc. Different policies have different impacts on China-US technology, and it is difficult to comprehensively assess the impact of China-US technological friction from a single policy change. Second, while the policy direction of ``small yard, high fence" may be clear for the US government, many details and the logic behind the policies remain hidden and are not articulated in official public policy or discourse. Some policy impacts arise from expectations about future policies, making it difficult to evaluate the overall impact of China-US technological friction simply by aggregating individual policies. The heterogeneous impacts on technological knowledge flow identified in this paper will be a compound effect of various overlapping policies, rather than a single effect. Therefore, this paper uses machine learning-based econometric methods, driven by data, to identify the effects of China-US technological friction on knowledge flow between the two countries.

\section{Data Description and Variable Selection}
\subsection{Data Description}
The data used in this paper mainly comes from three sources:
\textbf{Patent Application Data} from the China National Intellectual Property Administration (CNIPA) for invention patents in China.
\textbf{Invention Patent Citation Data} from the seven major patent-granting countries globally, as well as data on Chinese applicants’ invention patent applications at the United States Patent and Trademark Office (USPTO), the European Patent Office (EPO), and the World Intellectual Property Organization (WIPO), sourced from the Google Patents database.\footnote{The seven countries in question are China, the US, Japan, South Korea, Germany, the UK, and France. According to statistics from the World Intellectual Property Organization (WIPO), the total number of patents granted in these seven countries exceeds 95\% of the global total.}
\textbf{Citation Data} for US Invention Patents to Basic Science Papers, compiled from the public database on ``Reliance on Science" by \cite{marx2020reliance}.\footnote{Due to reasons such as initial reviews and information entry, there is a certain time lag in the publication of invention patent applications. Empirical studies show that, for observed patents granted, the average time from application to grant is about 2 years (e.g., \citealp{hall2005nber,seru2014firm,cornaggia2015does}). Some applications take far longer than 2 years to review, so the currently available granted data may not capture this delay. To fully reflect the volume of invention applications, we used invention publication data spanning a longer time period to construct the database, specifically using invention patent announcement data from 2015 to March 2023, in order to ensure that truncation errors due to publication lag are minimized \citep{lerner2017use}.} 

Based on these data sources, this paper constructs a multinational invention patent panel database with monthly frequency for China from 2015 to 2020.The database records monthly patent applications from China and international applications to China in 39,076 technology fields defined by international patent classification codes. To ensure that there is sufficient variation in patent application volumes for statistical analysis across different technological fields, we only used those fields with enough applications at the monthly level. The total number of observations for foreign invention patents granted to China is 5.26 million, while the number of observations for Chinese invention patent applications abroad is 1.55 million. Compared to existing literature, this database is the most granular and comprehensive microscopic database on knowledge flows in China’s various technology fields.

Additionally, the constructed panel database is matched with the basic science paper citation data from US patents from 1949 to 2018 compiled by \cite{marx2020reliance}. A total of 28,177 technology fields with citations to basic science papers were successfully matched. Finally, the panel database is matched with the patent citation data from the seven major technology countries, calculating the external citation volume of patents in 25,641 technology fields for both China and the US.

\subsection{Variable Selection}
\subsubsection{Cross-border invention patent applications and technological knowledge flow}
Cross-national patent applications are the direct vehicle for international knowledge flow \citep{eaton2002technology} and are therefore the core variable for measuring cross-national technological knowledge flow. This paper focuses on two aspects of technological knowledge flow related to China: technological knowledge flow from other countries/regions (such as the US) to China, and technological knowledge flow from China to other countries/regions (such as the US). The former is measured by foreign invention patent applications to China, and the latter by Chinese invention patent applications abroad. This paper specifically examines the impact of Sino-US technological friction on the inflow of US and non-US patents into China, as well as the effect of Chinese patent applications at the US Patent and Trademark Office (USPTO), the European Patent Office (EPO), and WIPO.

\subsubsection{Basic Science Dependence }
Basic science dependence is the first core explanatory variable for studying the impact of Sino-US technological friction. We aggregate the weighted citation volume of US invention patents to basic science papers in various technology fields, as compiled by \cite{marx2020reliance}.\footnote{The weights here refer to those used by \cite{marx2020reliance} to evaluate the academic level of citation journals.} This variable reflects the importance of basic research in each technology field.

\subsubsection{US Technological Strength Distribution Across Technology Fields}
The technological strength of the US in various fields is the second core explanatory variable for studying the impact of Sino-US technological friction. Using patent grants at the US Patent and Trademark Office from 1990 to 2017, this paper calculates the share of US patent grants in each technology field to measure the distribution of US technological strength.

\subsection{Sino-US Technological Gap}
The Sino-US technological gap is the third key explanatory variable for studying factors influencing Sino-US technological friction. Based on patent citation data from the seven major patent-granting countries, we calculate the external citation volume for China and the US in each IPC field.\footnote{External citations refer to citations made by non-domestic patents and are commonly used to correct for the preferences of inventors or patent office examiners regarding domestic patent citations. To ensure the comparability of patent citation counts, we used external citations excluding citations of domestic patents when calculating citation counts for both China and the US across the seven major granting countries.} The following Equation (\ref{eq:gap}) is used to measure the technological distance of China relative to the technological boundary of the US in each field:

\begin{equation}\label{eq:gap}
    gap_i = \frac{UScit_i}{UScit_i + CNcit_i},
\end{equation}
where, $UScit_i$ represents the number of citations to US patents in technology field $i$ by patents from other countries; $CNcit_i$ represents the number of citations to Chinese patents in technology field $i$ by patents from other countries. The larger the value of $gap_i$, the greater the technological gap between China and the US in that field. Table~\ref{tab:desc_stat} provides descriptive statistics for the variables used in this study.

\begin{table}[htbp]
\centering
\caption{Descriptive statistics}
\label{tab:desc_stat}
\begin{threeparttable}
\begin{tabular}{lcccc}
\toprule
\textbf{Variable Name} & \textbf{Data Source} & \textbf{Sample Size} & \textbf{Mean} &\textbf{Std. Dev.} \\
\midrule
\textbf{Number of patent applications} & & & &\\
\quad \textit{A. Patent Application Entry} & CNIPA & & & \\
\quad\quad US & & 1,103,868 & 0.627 & 1.805 \\
\quad\quad Non-US Foreign & & 1,749,960 & 1.187 & 3.347 \\
\quad\quad Mainland China & & 2,411,772 & 7.979 & 33.242 \\
\quad\quad Total & & 5,265,600 & 4.181 & 22.864 \\
\quad \textit{B. Patent Application Output} & Google Patents & & & \\
\quad\quad USPTO & & 556,800 & 0.778 & 3.013 \\
\quad\quad EPO & & 358,236 & 0.568 & 1.900 \\
\quad\quad WIPO & & 637,110 & 0.822 & 3.179 \\
\quad\quad Total & & 1,552,146 & 0.748 & 2.872 \\
Reliance on Basic Science & Marx \& Fuegi (2020) & 25,641 & 355.873 & 294.088 \\
Distribution of US Technological\\ Strength (Base Point) & Google Patents & 25,641 & 0.39 & 1.510 \\
Technological Gap between\\China and US & Google Patents & 25,641 & 0.975 & 0.070 \\
\bottomrule
\end{tabular}
\end{threeparttable}
\end{table}

\section{C-Lasso ML Method and Model Specification}

To identify the heterogeneous effects of Sino-US technological friction, this paper introduces the machine learning method based on the Lasso model proposed by \cite{su2016identifying} to conduct event study analysis of the heterogeneous policy effects of Sino-US technological friction. \cite{su2016identifying} use this method to estimate the heterogeneous coefficients of panel regressions by applying a penalty function that converges the individual regression coefficients to finite group coefficients, allowing for data-driven identification of groups affected by different policies.

\subsection{C-Lasso ML Method and Group Heterogeneous with Panel Data}

\cite{su2016identifying} proposed a machine learning method based on penalty functions to identify potential group structures in panel data. The following individual heterogeneity panel regression structure is considered:
\begin{equation}
    y_{it} = x_{it}'\beta_i + \gamma_i + \varepsilon_{it},
\end{equation}
where $i$ is the individual index and $t$ is the time index. $\beta_i$ captures the individual-specific slope coefficients. When $\beta_i = \beta$, this model degenerates into a homogeneous fixed-effects model. Many economic theories suggest that the choices of economic agents are not entirely homogeneous, but heterogeneous, such as ``club convergence" in cross-national economic growth studies or multiple equilibria in industrial organization theory. The method for addressing this heterogeneity in econometrics is applied to such issues. In the framework of \cite{su2016identifying}, $\beta_i$ is group-specific, i.e., $\beta_i = \beta_k$, where $i \in \{1, 2, ..., N\}$ and $k \in \{1,2, ..., K\}$, with $K \leq N$.

There are $K$ groups and individuals within the same group are homogeneous, while groups differ in terms of $\beta_i$. \cite{su2016identifying} found strong parameter sparsity in heterogeneous panel regressions, which aligns well with the Lasso method in machine learning. Based on this, \cite{su2016identifying} proposed the Classifier-Lasso (C-Lasso) method to identify heterogeneous group coefficients in panel regressions given a fixed number of groups.

We focus on the linear model of C-Lasso \citep{lee2022lasso} with the group penalty least squares (PLS) method. Given an adjustable parameter c\footnote{The adjustable parameter c reflects the relative importance of the penalty function term compared to the least squares method. When c = 0, PLS degenerates to OLS.} and group number K, the PLS estimator is the solution to the following optimization problem:

\begin{equation}
\underset{\{\beta_i\}_{i=1}^{N}, \{\alpha_k\}_{k=1}^{K}}{min} 
\underbrace{\frac{1}{NT}\sum_{i=1}^{N}\sum_{t=1}^{T}(y_{it}-x'_{it}\beta_i)^2}_{\textit{Least Squares Method}}
+ \underbrace{\frac{c\,Var(y_{it})}{N\,T^{1/3}}\sum_{i=1}^{N}\sum_{k=1}^{K}\|\beta_i - \alpha_k\|^2}_{\textit{Penalty Term}}.
\end{equation}

The penalty term ensures that the slope coefficients $\beta_i$ converge to the group slope coefficients $\alpha_k$, and the optimization problem can be solved using the iterative algorithm constructed by \cite{su2016identifying}.

\subsection{Model Specification}

\subsubsection{Identifying Heterogeneous Groups Using C-Lasso}
This paper combines the event study method with C-Lasso to identify IPC groups that were negatively impacted by the policy, i.e., the ``small yard, high fence". After C-Lasso grouping, the paper constructs a triple-differencing (DDD) framework to empirically analyze the treatment and control groups identified by the data-driven approach. The validity of the C-Lasso method is assessed through parallel trend tests and placebo tests.
The event study model with heterogeneous group parameters is specified as:

\begin{equation}\label{eq:uss}
\begin{aligned}
\ln(1 + USSq_{it}) &= \beta_{1,k}Post_t + \beta_{2,k}\ln(1 + CounterSq_{it}) + \beta_{3,k}Trend_{it} + a_i \times m_t + \varepsilon_{it}, \\
&\forall i \in group_k, \quad k \in \{1, 2, ..., K\},
\end{aligned}
\end{equation}

where $i$ represents technological fields, $t$ represents months, and $k$ represents heterogeneous groups. The dependent variable, $USSq_{it}$, represents US patent applications in China in technology field $i$, or Chinese patent applications in the US in the knowledge-outflow regression. The control variable, $CounterSq_{it}$, represents patent applications in China by non-US countries in the knowledge-inflow. The variable $Trend_{it}$ is a time trend to account for any underlying trends in the data. Additionally, the interaction term of technological field fixed effects $a_i$ and month fixed effects $m_t$ is added to control for other effects that remain unchanged within each month of the year within the technological field. $\varepsilon_{it}$ is the error term.

The estimated coefficients $\beta_{1,k}$ indicate the impact of Sino-US technological friction on patent application inflows and outflows within different technology groups. The key identification strategy is the dummy variable $Post_t$ which captures the treatment effect after the escalation of Sino-US technological tensions. The time point of the event is chosen as January 2018, when the US first imposed substantive tariff sanctions.

The model also includes group-specific fixed effects to control for heterogeneity across groups. The variable $group_k$ represents the index set of the heterogeneous group $k$ divided by the C-Lasso method. In the parameter selection, this paper sets the maximum number of groups $K$ to 3, and the importance of the penalty function term $c$ to 0.25.\footnote{ For robustness, we tried multiple combinations of (K, c) and found that the identification results for heterogeneous groups were highly robust to changes in the values of (K, c). Due to space constraints, the main text only shows the identification results for the combination (3, 0.25).} Based on the heterogeneous groups identified by C-Lasso, the paper classifies technology fields that are negatively impacted by the policy as the treatment group, as shown in Equation (\ref{eq:countyard}):

\begin{equation}\label{eq:countyard}
Courtyard_i = 1(i \in \{group_k|\beta_{1,k} < 0\}).
\end{equation}

\subsubsection{Event Study}

Based on the heterogeneous group results identified by the C-Lasso model, the paper establishes the following triple-differencing (DDD) model to examine the causal effects of the ``small yard, high fence" policy. The standard triple-differencing model is shown in Equation (\ref{eq:Sq}):

\begin{equation}\label{eq:Sq}
Sq_{irt} = \left(\beta^{DDD}Post_t \times Courtyard_i \times US_r 
+ \gamma Year_t \times Courtyard_i \times Region_r + a_{ir} \times m_t + \varepsilon_{irt}\right),
\end{equation}

where $i$ represents technological fields, $t$ represents months, and $r$ represents the source or destination country. The dependent variable $Sq_{irt}$ represents the application volume. The double-group variable $Courtyard_i \times US_r$ represents US ``small yard, high fence" technology fields. The term $Year_t \times Courtyard_i \times Region_r$ controls the annual trend of applications in ``small yard, high fence" technology fields across countries. By adding the interaction term of country-specific technology field fixed effects $a_{ir}$ and monthly calendar effects $m_t$, the model controls for other effects that are constant within each technology field across months. $\varepsilon_{irt}$ is the error term.

Similarly, this paper uses an incremental triple-differencing model to conduct an event study for 20 months before and after the ``small yard, high fence" policy and a pre-policy parallel trend test, as shown in Equation (\ref{eq:sqirt}):

\begin{equation}\label{eq:sqirt}
\begin{aligned}
Sq_{irt} &= \sum_{s=-20, s \neq 0}^{20}\left(\beta_s^{DDD}Period_s \times Courtyard_i \times US_r + \beta_s^{DD,1}Period_s \times Courtyard_i \right. \\
&\quad\left.+ \beta_s^{DD,2}Period_s \times Region_r + \tau_s Period_s\right) + \gamma Year_t \times Courtyard_i \times Region_r + a_{ir} \times m_t + \varepsilon_{irt}.
\end{aligned}
\end{equation}

\subsection{Determinants of ``Small yard, High fence" Policy}

Based on the heterogeneous group results identified by the C-Lasso model, the paper establishes the following unit probability (Probit) model in Equation (\ref{eq:probit_model}) to examine the factors determining the ``small yard, high fence" policy from three aspects: basic science dependence, US technological strength distribution, and the technological gap between China and the US:

\begin{equation}\label{eq:probit_model}
\text{Pr}[Courtyard_i = 1 | X] = \Phi(\alpha + \beta_1 ScienceCit_i + \beta_2 USshare_i + \beta_3 gap_i),
\end{equation}

where the dependent variable represents the probability that technology field $i$ is negatively impacted by the policy. The variable $ScienceCit_i$ represents the science dependence of technology field $i$; $USshare_i$ represents the share of US patent grants in technological field $i$; and $gap_i$ represents the technological gap between China and the US in technology field $i$, as measured by Equation (\ref{eq:gap}).

\section{Heterogeneous Groups Identification}

\subsection{Heterogeneous Groups Identification Results}
Table \ref{tb:H-identification} presents the C-Lasso model in Equation (\ref{eq:uss}) results for identifying heterogeneous groups, with Panel A and Panel B showing the regression results for patent inflows and outflows, respectively.

\begin{table}[htbp]\centering
\caption{Heterogeneous group identification results}
\label{tb:H-identification}
\resizebox{\textwidth}{!}{%
\begin{tabular}{lccccccc}
\hline
 & \multicolumn{7}{c}{Panel A: Patent Application Inflow} \\
\cline{2-8}
Dependent Variable: \textit{USSq} & Pooled & \multicolumn{3}{c}{C-Lasso} & \multicolumn{3}{c}{Post-Lasso} \\
(US Patent Applications in China) & Regression & \multicolumn{3}{c}{} & \multicolumn{3}{c}{} \\
 & (A1) & (A2) & (A3) & (A4) & (A5) & (A6) & (A7) \\
 &  & Group1 & Group2 & Group3 & Group1 & Group2 & Group3 \\
\hline
Post & -0.053*** & -0.662 & -0.098 & 0.498 & -0.662*** & -0.099*** & 0.498*** \\
 & (0.002) &  &  &  & (0.005) & (0.003) & (0.004) \\
\textit{CounterSq} & 0.022*** & 0.0185 & 0.0299 & 0.0126 & 0.019*** & 0.030*** & 0.013*** \\
(Non-US Foreign Applications) & (0.001) &  &  &  & (0.001) & (0.001) & (0.001) \\
\textit{Trend} & 0.002*** & 0.0144 & 0.003 & -0.0099 & 0.014*** & 0.003*** & -0.010*** \\
 & (5e-5) &  &  &  & (0.0001) & (0.0001) & (0.0001) \\
Fixed Effects: &  &  &  &  &  &  &  \\
Country $\times$ Technology Field $\times$ Month & Y & Y & Y & Y & Y & Y & Y \\
\hline
 & \multicolumn{7}{c}{Panel B: Patent Application Outflow} \\
\cline{2-8}
Dependent Variable: \textit{USSq} & Pooled & \multicolumn{3}{c}{C-Lasso} & \multicolumn{3}{c}{Post-Lasso} \\
(Chinese Patent Applications in the US) & Regression & \multicolumn{3}{c}{} & \multicolumn{3}{c}{} \\
 & (B1) & (B2) & (B3) & (B4) & (B5) & (B6) & (B7) \\
 &  & Group1 & Group2 & Group3 & Group1 & Group2 & Group3 \\
\hline
Post & -0.006 & -0.453 & 0.0005 & 0.427 & -0.454*** & 0.001 & 0.428*** \\
 & (0.004) &  &  &  & (0.006) & (0.006) & (0.006) \\
\textit{CounterSq} & 0.187*** & 0.0528 & 0.57 & 0.055 & 0.053*** & 0.570*** & 0.055*** \\
(Chinese Patent Applications in Europe) & (0.002) &  &  &  & (0.002) & (0.004) & (0.002) \\
\textit{Trend} & 0.004*** & 0.013 & 0.002 &- 0.0043 & 0.013*** & 0.002*** & -0.004*** \\
 & (8e-5) &  &  &  & (0.0001) & (0.0001) & (0.0001) \\
Fixed Effects: &  &  &  &  &  &  &  \\
Country $\times$ Technology Field $\times$ Month & Y & Y & Y & Y & Y & Y & Y \\
\hline
\multicolumn{8}{l}{\footnotesize{Notes: (1) Robust standard errors clustered at the country $\times$ technology field $\times$ month level are in parentheses.}} \\
\multicolumn{8}{l}{\footnotesize{(2) *, **, and *** denote significance at the 10\%, 5\%, and 1\% levels, respectively.}} \\
\end{tabular}%
}
\end{table}

In the A1 and B1 columns, the homogeneous mixed regression results without grouping are displayed. The mixed regression results indicate that, overall, Sino-US technological friction has had little substantive impact on the flow of technological knowledge between China and the US For US patent applications to China, the decrease is only 5\%, while China’s patent applications in the US were barely affected. Columns A2-A4 and B2-B4 show the regression results for knowledge inflows and outflows identified by the C-Lasso model, with the second to fourth columns corresponding to the heterogeneous regression parameters for the three groups. The table shows that the parameters in the second column correspond to technology field groups significantly negatively impacted by Sino-US technological friction. From the perspective of technological knowledge inflows, 4,633 IPC-coded fields were significantly negatively impacted by the policy, accounting for 11.8\% of the total IPC-coded fields, while 3,351 IPC-coded fields were significantly negatively impacted by the policy from the perspective of technological knowledge outflows, accounting for 8.6\%. A total of 1,372 IPC-coded fields were significantly impacted by the policy in both inflow and outflow dimensions, accounting for 3.5\% of the total IPC-coded fields .\footnote{The estimated model for the Post-Lasso estimator after grouping is:

$\ln(1+USSq_{it})=(\beta_{1,k}Post_t+\beta_{2,k}\ln(1+CounterSq_{it})+\beta_{3,k}Trend_{it})\times group_k+a_i\times m_t+\varepsilon_{it},$
where \textit{group$_k$} is the group dummy variable identified by C-Lasso.}

Using the group results from the C-Lasso model, we conducted group-based regression, and the results in A5-A7 and B5-B7 demonstrate the significance of the findings. The regression results reveal the core conclusion of this paper: the US technological friction strategy towards China overall represents a ``small yard, high fence" policy rather than a comprehensive ``decoupling". Whether from the perspective of technological knowledge inflows or outflows, Sino-US technological friction has not led to a dramatic change in total patent applications. Simple event study methods can easily obscure the true impact of Sino-US technological friction on technological knowledge flows. After identifying heterogeneous groups with the C-Lasso model, we find that Sino-US technological friction has a strong heterogeneous impact on technological knowledge flows between China and the US In technology fields negatively impacted by the policy, US patent applications to China dropped by 66\%, while China’s patent applications to the US fell by 45\%, representing a substantial decrease in Sino-US technological knowledge flow.

\subsection{Characteristics of Technology Fields Negatively Affected by the Policy}
To better understand the characteristics of the technology fields negatively affected by the policy in qualitative terms, we match the affected technology fields to the industrial level. Based on the ``Strategic Emerging Industry Classification and International Patent Classification Reference Table (2021)"\footnote{Source: \url{https://www.cnipa.gov.cn/art/2021/2/10/art_75_156716.html}} from the China National Intellectual Property Administration, we match all IPC-coded technology fields to the 40 secondary classifications of strategic emerging industries in the ``13th Five-Year Plan for the Development of Strategic Emerging Industries".
Based on this, the paper establishes the following Equation (\ref{eq:treatratio}) to measure the impact of the ``small yard, high fence" policy on strategic emerging industries:

\begin{equation}
\label{eq:treatratio}
TreatRatio_i = 100 \times \left(\frac{YardShare_i}{GrossShare_i} - 1\right),
\end{equation}

where $YardShare_i$ represents the ratio of the total patent applications from technology fields within strategic emerging industry $i$ that are negatively impacted by the policy to the total patent applications in the affected technology fields; $GrossShare_i$ represents the share of the total patent applications from technology fields within strategic emerging industry $i$ in the overall total patent applications; $TreatRatio_i$ reflects the relative intensity of the ``small yard, high fence" policy's impact on strategic emerging industries. The higher the $TreatRatio_i$ the more significant the policy's impact on industry $i$ If the intensity of the policy's impact on a particular strategic emerging industry is equal to the average impact across all industries, then $TreatRatio_i=0.$

Figure \ref{fig:degree_emerging_industry} shows the distribution of $TreatRatio_i$ for 40 strategic emerging industries in terms of both patent inflows and outflows. It can be observed that the ``small yard, high fence" policy has a significant heterogeneous impact on both the inflow and outflow dimensions of technological knowledge.

\subsubsection{Technological Inflow Dimension}
\begin{figure}
\centering
\includegraphics[width=1\textwidth]{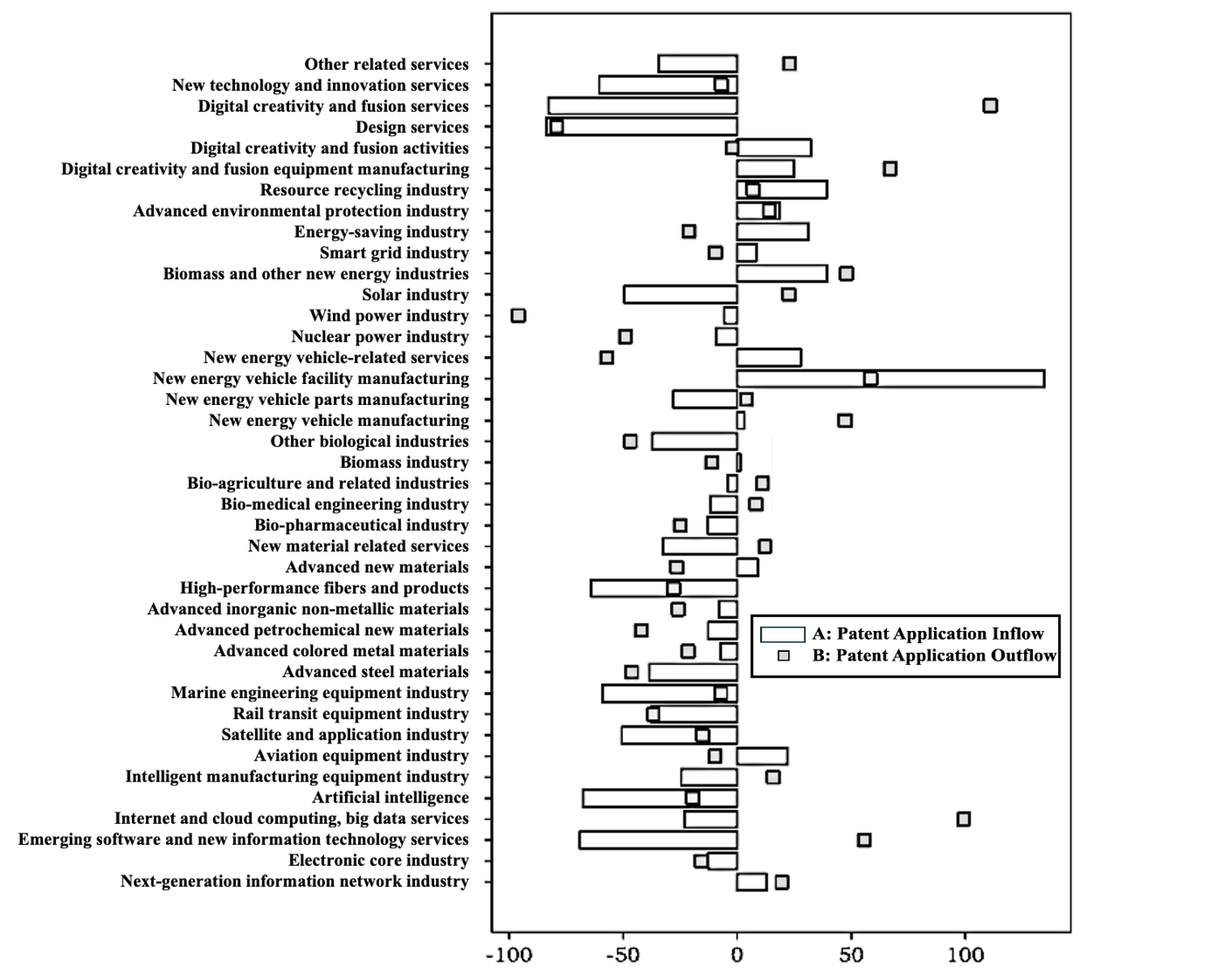}
\caption{The ``small yard, high fence" degree of strategic emerging industries in two dimensions}
\label{fig:degree_emerging_industry}
\end{figure}

As shown in Figure \ref{fig:degree_emerging_industry}, in the dimension of US patent applications to China, the ``small yard, high fence" industries are relatively concentrated. Firstly, in the energy-saving and environmental protection industry related to new energy, all sub-industries except for solar, wind, and nuclear power have been widely and significantly negatively impacted, with the most significant impact on the biomass energy and other new energy industries. Secondly, in the new energy vehicle industry, the impact of Sino-US technological friction on technological diffusion mainly affects the sub-industries of new energy vehicle-related services and equipment manufacturing, with the new energy vehicle-related equipment manufacturing industry being the most affected in terms of technological inflow. Finally, industries like digital culture and creativity, aviation equipment, and next-generation information networks have also been notably negatively impacted.

\subsubsection{Technological Outflow Dimension}
Compared to patent applications from the US to China, the industries affected by patent applications from China to the US are more widespread. First, similar to the technology diffusion inflow dimension, sectors like bio-intelligence, other new energy industries, new energy vehicle-related equipment manufacturing, digital creative technology equipment manufacturing, and next-generation information network industries also experience negative impacts from China-US technological friction in terms of technological competition. Secondly, some industries are only affected in the technology diffusion outflow dimension, including rapidly developing Chinese sectors such as internet software and new energy vehicle manufacturing industries.

\section{Event Study and Robustness Tests}
In this section, we first use the heterogeneity group identification results discussed earlier and apply a triple-differencing framework to study the impact of China-US technological friction on knowledge flow between the two countries. Additionally, we verify the robustness of the conclusions through parallel trend tests and placebo tests.

\subsection{Benchmark Regression}
Table \ref{tb:benchmark} presents the estimated results of the baseline regression model in Equation (\ref{eq:Sq}), using three types of models: Ordinary Least Squares (OLS), Poisson Pseudo Maximum Likelihood (PPML), and Tobit model to estimate the standard difference-in-differences framework. The standard triple-differencing mehtod identifies the causal effect of US technological restrictions on China by examining the application volume changes before and after the policy for technology fields affected by the policy. The empirical results show that the heterogeneous policy effects have led to a significant decrease in US patent applications in the affected fields, ranging from 12\% to 90\% compared to the control group (non-US regions). Similarly, Chinese patent applications in the US decreased by 9\% to 40\% compared to the control group (European Patent Office and World Intellectual Property Organization). Quantitatively, technology diffusion in the inflow dimension is more strongly affected by the ``small yard, high fence" policy than in the outflow dimension. Based on the annual trend coefficients of US applications in the affected technology fields, we observe that knowledge flow between China and the US in these sectors is rapidly growing. Therefore, the affected technology fields are also the focal points of China-US technological interaction, and omitting this variable leads to an underestimate of the causal effects. This will be reaffirmed in subsequent multi-period event studies.

\begin{table}[htbp]\centering
\caption{Benchmark regression for the standard triple-differencing model}
\label{tb:benchmark}
\resizebox{\textwidth}{!}{%
\begin{tabular}{lccccc}
\hline
 & \multicolumn{5}{c}{Panel A: Patent Application Inflow} \\
\cline{2-6}
Dependent Variable: $Sq$ & (A1) & (A2) & (A3) & (A4) & (A5) \\
(Patent Applications in China) & OLS & OLS & PPML & Tobit & Tobit \\
\hline
Post $\times$ Courtyard $\times$ US & -0.119*** & -0.438*** & -0.424*** & -0.240*** & -0.901*** \\
 & (0.002) & (0.004) & (0.013) & (0.006) & (0.008) \\
Year $\times$ Courtyard $\times$ US &  & 0.106*** &  &  & 0.223*** \\
 &  & (0.001) &  &  & (0.003) \\
Country Fixed Effects &  &  &  & Y & Y \\
Technology Fixed Effects &  &  &  & Y & Y \\
Month Fixed Effects &  &  &  & Y & Y \\
Country $\times$ Technology$\times$ Month & Y & Y & Y &  &  \\
Observations & 5265600 & 5265600 & 5265600 & 5265600 & 5265600 \\
R-sq/\textit{pseudo} R-sq & 0.774 & 0.5145 & 0.808 & 0.0607 & 0.0629 \\
\hline
 & \multicolumn{5}{c}{Panel B: Patent Application Outflow} \\
\cline{2-6}
Dependent Variable: $Sq$ & (B1) & (B2) & (B3) & (B4) & (B5) \\
(Chinese Patent Applications Overseas) & OLS & OLS & PPML & Tobit & Tobit \\
\hline
Post $\times$ Courtyard $\times$ US & -0.089*** & -0.359*** & -0.230*** & -0.100*** & -0.397*** \\
 & (0.003) & (0.004) & (0.018) & (0.006) & (0.006) \\
Year $\times$ Courtyard $\times$ US &  & 0.090*** &  &  & 0.101*** \\
 &  & (0.001) &  &  & (0.001) \\
Country Fixed Effects &  &  &  & Y & Y \\
Technology Fixed Effects &  &  &  & Y & Y \\
Month Fixed Effects &  &  &  & Y & Y \\
Country $\times$ Technology$\times$ Month & Y & Y & Y &  &  \\
Observations & 1552146 & 1552146 & 1552146 & 1552146 & 1552146 \\
R-sq/\textit{pseudo} R-sq & 0.492 & 0.494 & 0.480 & 0.024 & 0.0262 \\
\hline
\multicolumn{6}{l}{\footnotesize{Notes: (1) Robust standard errors clustered at the country$\times$technology field$\times$month level are in parentheses.}} \\
\multicolumn{6}{l}{\footnotesize{*, **, and *** indicate significance at the 10\%, 5\%, and 1\% levels, respectively. (3) Models (A/B1) and (A/B2) use}} \\
\multicolumn{6}{l}{\footnotesize{ordinary least squares (OLS) estimation, with the dependent variable log-transformed after adding 1. (4) Model (A/B3)}} \\
\multicolumn{6}{l}{\footnotesize{uses Poisson pseudo-maximum likelihood (PPML) estimation. (5) Models (A/B4) and (A/B5) use Tobit estimation}} \\
\multicolumn{6}{l}{\footnotesize{with sample merging, where the dependent variable is log-transformed after adding 1, and the left censoring point is}} \\
\multicolumn{6}{l}{\footnotesize{0. The reported coefficients are average marginal effect estimates for E[ln(1+$sq_{it}$)|X].}} \\
\end{tabular}%
}
\end{table}

Table \ref{tb:did} decomposes the baseline regression into different source/destination country differences before and after the policy. In the technology diffusion inflow dimension, we examine the application rate changes in Chinese patent offices from the US, non-US international regions, and local China. In the technology diffusion outflow dimension, we study changes in Chinese patent applications in US patent offices, European Patent Office, and WIPO. In the inflow dimension, the difference-in-differences results show that only US applications in China experienced a significant decline due to the ``small yard, high fence" policy, with a reduction of 12\% to 40\% compared to non-policy affected fields. In the outflow dimension, the results show a decline in Chinese applications abroad, including in the US, European Patent Office, and WIPO, by 5\% to 40\%, and 0.6\% to 20\%, respectively. We believe this spillover effect of US technological restrictions may stem from the shrinking US market, which reduced the innovation incentives for Chinese technology companies primarily focused on the US market, thus lowering their patent filings in all overseas markets.

\begin{table}[htbp]\centering
\caption{Difference-in-differences decomposition of baseline regression}
\label{tb:did}
\resizebox{\textwidth}{!}{%
\begin{tabular}{lcccccc}
\hline
 & \multicolumn{6}{c}{Panel A: Patent Application Inflow} \\
\cline{2-7}
 & \multicolumn{3}{c}{OLS} & \multicolumn{3}{c}{PPML} \\
Dependent Variable: $Sq$ & (A1) & (A2) & (A3) & (A4) & (A5) & (A6) \\
(Patent Applications in China) & US & Non-US & China & US & Non-US & China \\
\hline
Post $\times$ Courtyard & -0.119*** & -0.001 & 0.005** & -0.396*** & -0.027*** & 0.035*** \\
 & (0.002) & (0.002) & (0.002) & (0.009) & (0.007) & (0.011) \\
Technology$\times$ Month & Y & Y & Y & Y & Y & Y \\
Observations & 1103868 & 1749960 & 2411772 & 1103868 & 1749960 & 2411772 \\
R-sq/\textit{pseudo} R-sq & 0.456 & 0.605 & 0.789 & 0.384 & 0.525 & 0.812 \\
\hline
 & \multicolumn{6}{c}{Panel B: Patent Application Outflow} \\
\cline{2-7}
 & \multicolumn{3}{c}{OLS} & \multicolumn{3}{c}{PPML} \\
Dependent Variable: $Sq$ & (B1) & (B2) & (B3) & (B4) & (B5) & (B6) \\
(Chinese Patent Applications Overseas) & US & Europe & Global & US & Europe & Global \\
\hline
Post $\times$ Courtyard & -0.051*** & -0.006*** & -0.014*** & -0.405*** & -0.129*** & -0.200*** \\
 & (0.001) & (0.001) & (0.001) & (0.014) & (0.016) & (0.015) \\
Technology$\times$ Month & Y & Y & Y & Y & Y & Y \\
Observations & 1103868 & 1749960 & 2411772 & 1103868 & 1749960 & 2411772 \\
R-sq/\textit{pseudo} R-sq & 0.605 & 0.475 & 0.579 & 0.516 & 0.358 & 0.498 \\
\hline
\multicolumn{7}{l}{\footnotesize{Notes: (1) Robust standard errors clustered at the country$\times$technology field$\times$month level are in parentheses.}} \\
\multicolumn{7}{l}{\footnotesize{(2) *, **, and *** indicate significance at the 10\%, 5\%, and 1\% levels, respectively. (3) Models (A/B1)--(A/B3) use}} \\
\multicolumn{7}{l}{\footnotesize{ordinary least squares (OLS) estimation, with the dependent variable log-transformed after adding 1. (4) Models}} \\
\multicolumn{7}{l}{\footnotesize{(A/B4)--(A/B6) use Poisson pseudo-maximum likelihood (PPML) estimation.}} \\
\end{tabular}%
}
\end{table}

\subsection{Benchmark Regression}
\subsubsection{Event Study and Parallel Trends Test }

Using the event study framework defined in Equation (\ref{eq:sqirt}), we conduct an event study and parallel trend test on the impact of US tariffs on China over the 20 months before and after implementation.

\begin{figure}
\centering
\includegraphics[width=1\textwidth]{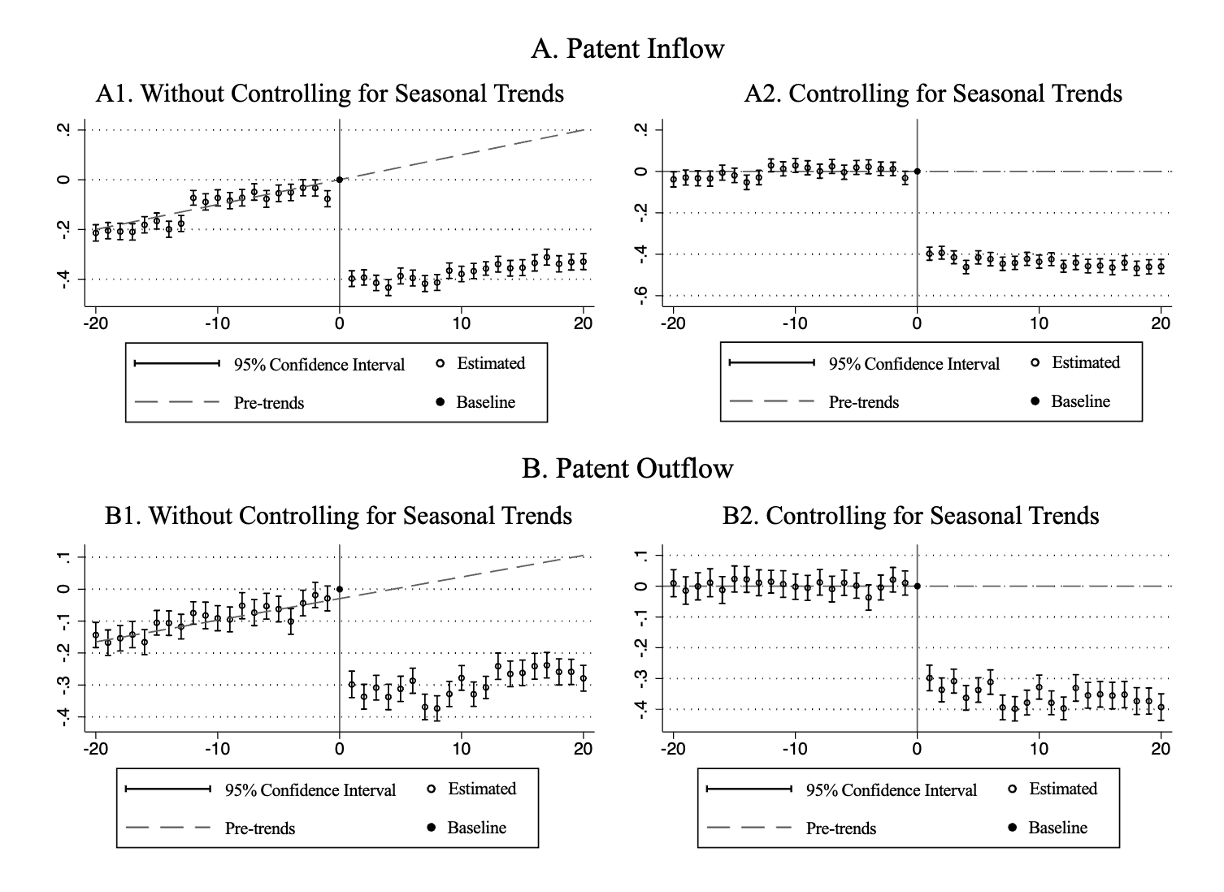}
\caption{The ``small yard, high fence" degree of strategic emerging industries in two dimensions}
\label{fig:event_study}
\end{figure}

Figure \ref{fig:event_study} reports the results of the OLS model estimated for model in Equation (\ref{eq:sqirt}), with Figures 2A and 2B showing the estimated results with and without controlling for annual time trends. As observed in the baseline regression, the affected technology fields show a sustained growth trend in both dimensions of technology diffusion compared to other source/destination countries. This annual trend is stable and significantly linear.

After controlling for the pre-treatment trend, the two subgraphs in Figure 2 show that the triple-differencing model satisfies the parallel trend assumption. In the US inflow dimension, the policy effect primarily presents as a level effect. After the actual imposition of tariffs, the patent application rate of the US in China for the affected fields decreased by 40\% compared to other sources. This trend remained highly persistent over the 20-month period studied. In the outflow dimension, the study finds that in addition to the level effect, there is also a certain degree of trend effect. After the tariffs, China’s patent applications in the US decreased by 30\%, and further dropped by 10\% in the next 20 months, even after controlling for the annual time trend effect.

\begin{figure}
\centering
\includegraphics[width=1\textwidth]{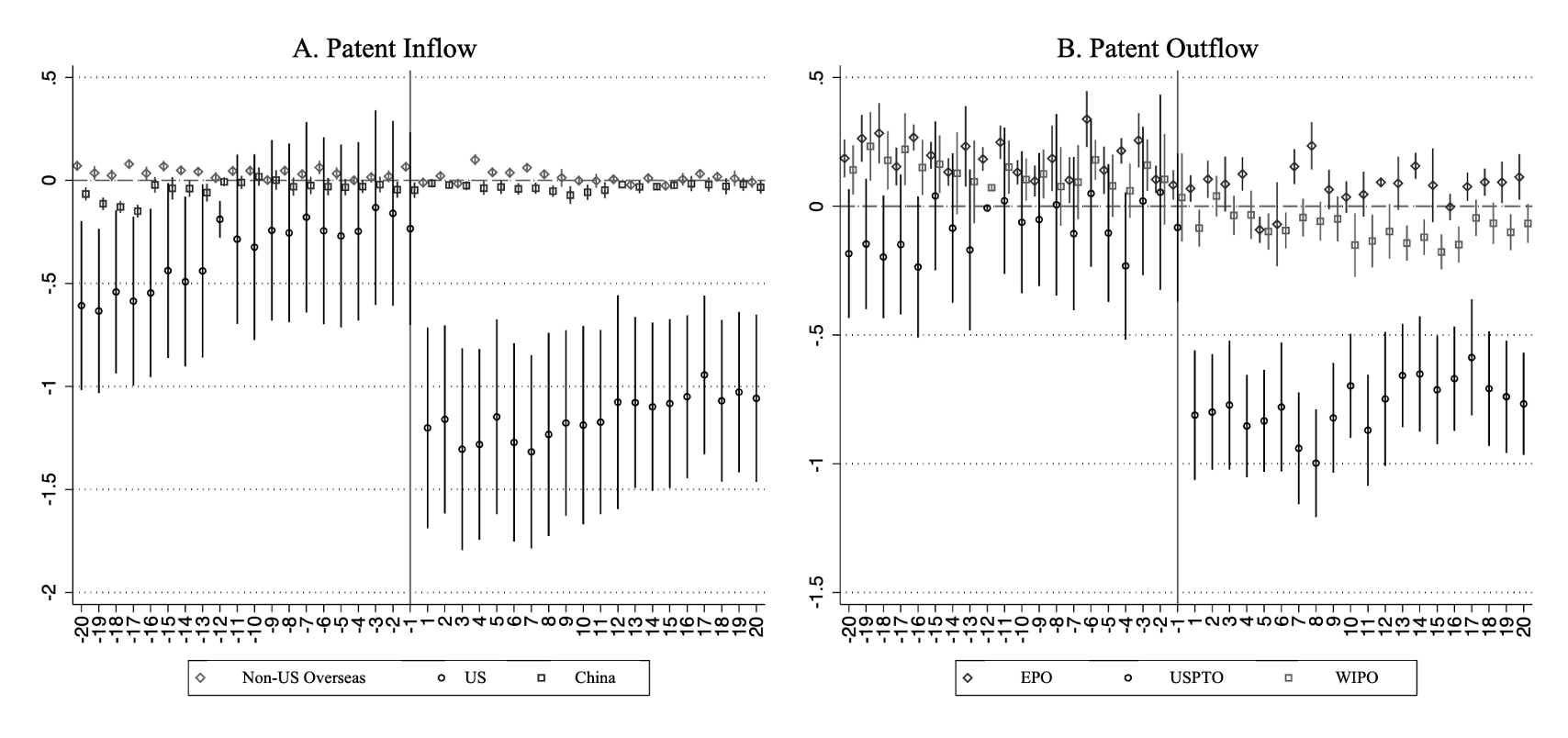}
\caption{The ``small yard, high fence" degree of strategic emerging industries in two dimensions}
\label{fig:progressive}
\end{figure}

Figure \ref{fig:progressive} reports the results of the double-difference decomposition of the event study model in Equation (\ref{eq:sqirt}) based on the PPML model. Similar to Table 4, we decompose the triple-differencing model results into progressive difference forms based on different source/destination countries around the policy implementation. From the decomposed progressive double-difference, we observe that the impact on China-US knowledge flow mainly comes from the changes in US applications. The applications from other source/destination countries in the ``small yard, high fence" fields showed almost no variation, confirming that the triple-differencing model results are not driven by trends in knowledge flow between China and other countries.

\subsubsection{Placebo Test}
\begin{figure}
\centering
\includegraphics[width=1\textwidth]{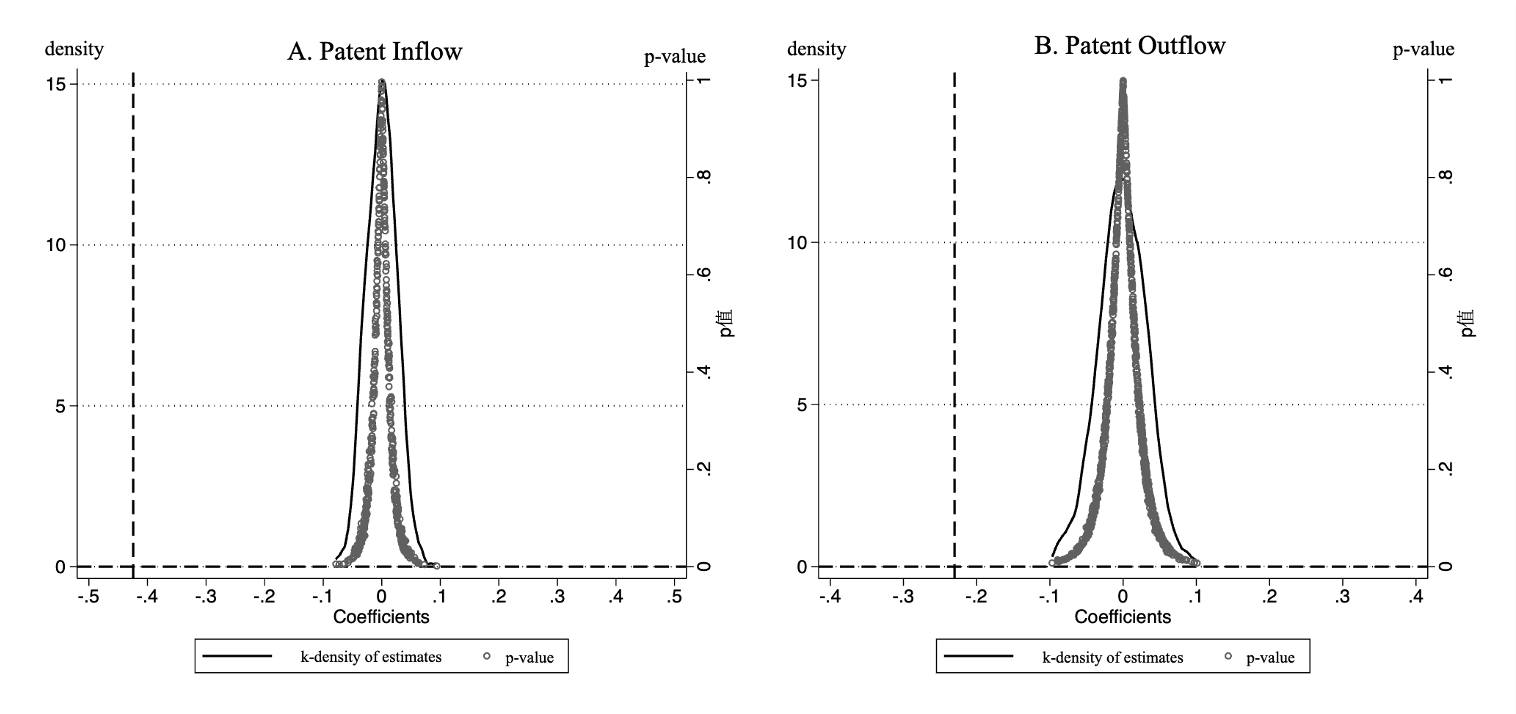}
\caption{The ``small yard, high fence" degree of strategic emerging industries in two dimensions}
\label{fig:placebo_test}
\end{figure}

In addition to the parallel trend test, we also conduct robustness checks on the baseline regression by repeatedly randomizing the treatment groups. We randomly select the same number of IPC code groups as placebo treatment fields to obtain placebo estimates. Due to space limitations, we report only the PPML model estimates for the baseline regression, but the results from OLS and Tobit models are consistent. The placebo estimation distribution, after 1,000 randomizations, is shown in Figure \ref{fig:placebo_test} and approximates a normal distribution with a mean of zero. The estimated coefficients from the baseline regression clearly stand out as outliers in the placebo distribution, confirming that the results are robust.

\section{Determinants of the Impacted Fields in China and the US}
The Probit model defined in Equation (\ref{eq:countyard}) examines the determinants of whether technology fields are impacted by policy, considering three dimensions: reliance on basic scientific research, distribution of US technological strength, and the China-US technological gap. Table \ref{tb:determinants} reports the determinants of affected technology fields for both patent application inflow, outflow, and those simultaneously impacted in both dimensions. 

\begin{table}[htbp]\centering
\caption{Determinants of impacted filds by the ``small yard, high fence'' policy}
\label{tb:determinants}
\resizebox{\textwidth}{!}{%
\begin{tabular}{lcccccc}
\hline
Dependent Variable: & \multicolumn{2}{c}{Patent Application Inflow} & \multicolumn{2}{c}{Patent Application Outflow} & \multicolumn{2}{c}{Inflow and Outflow} \\
Pr[$Courtyard=1$] & (1) & (2) & (3) & (4) & (5) & (6) \\
\hline
$ScienceCit$ & 0.008*** & 0.012*** & 0.039*** & 0.044*** & 0.012*** & 0.044*** \\
 & (0.002) & (0.002) & (0.002) & (0.002) & (0.001) & (0.002) \\

$USshare$ & 0.012*** & 0.011*** & 0.010*** & 0.005*** & 0.003*** & 0.005*** \\
 & (0.002) & (0.002) & (0.004) & (0.001) & (0.001) & (0.001) \\

$gap$ & -0.070* & -0.068* & -0.148*** & -0.106*** & -0.051*** & -0.106*** \\
 & (0.037) & (0.038) & (0.030) & (0.031) & (0.015) & (0.031) \\
Technical Field Effect: IPC4 Code &  & Y &  & Y &  & Y \\
Observations & 25641 & 25641 & 25641 & 25641 & 25641 & 25641 \\
\hline
\multicolumn{7}{l}{\footnotesize{Notes: (1) Robust standard errors clustered at the IPC4 code level are in parentheses.}} \\
\multicolumn{7}{l}{\footnotesize{(2) *, **, and *** indicate significance at the 10\%, 5\%, and 1\% levels, respectively.}} \\
\multicolumn{7}{l}{\footnotesize{(3) All estimated coefficients in the table are average marginal effects.}} \\
\end{tabular}%
}
\end{table}

The reported coefficients are the average marginal effects, showing how different factors influence the probability of a technology field being affected by the policy at the margin. First, technology fields that are more dependent on basic scientific research are more likely to be impacted by China-US technological friction. For example, a 10\% increase in a field’s reliance on academic journal citations will increase the probability of it being affected by about 1\% for inflows, 4\% for outflows, and 1\% to 4\% for both dimensions. Secondly, fields with greater US technological strength are more likely to be impacted by the policy. Specifically, a 1 basis point (0.01\%) increase in the US’s share of patent grants in a field increases the probability of being affected by 1\% for inflows, 0.5\% to 1\% for outflows, and 0.4\% for both dimensions. Finally, the smaller the technological gap between China and the U.S., the more likely the field is to be affected. A 10\% reduction in the technological gap between China and the U.S. increases the probability of being affected by 0.7\% for inflows, 1\% to 1.5\% for outflows, and 0.5\% to 1\% for both.

\section{Conclusion}

This paper uses invention patent application data from China, the U.S., Europe, and the World Patent Organization, along with machine learning methods, to precisely identify the technology fields impacted by China-U.S. technological friction. In the groups of technology fields negatively affected by the policy, the ``small yard, high fence" technological restrictions have had a significant negative impact on China-U.S. technology flow. Based on the C-Lasso models identification of technology fields affected by the policy, this paper quantitatively studies the impact of the ``small yard, high fence" policy on China-U.S. technology flow using a triple-differencing framework. The empirical results show that, compared to the control group technology fields and the technology fields in other countries/regions, only the U.S. treatment group technology fields saw a significant decrease of about 44\% and 36\% in technology diffusion inflows and outflows, respectively. The progressive triple-differencing model results indicate that this level effect remains highly persistent for 20 months after the policy’s implementation. Furthermore, using the ``Reliance on Science" public database for U.S. invention patents compiled by Marx \& Fuegi (2020) and citation network data from seven major patent-granting countries globally, the paper investigates the factors influencing the U.S. ``small yard, high fence" technology field selection strategy. The empirical conclusion shows that technology fields more reliant on basic scientific research, those with more concentrated U.S. technological strength, and those with a smaller China-U.S. technological gap are more likely to be affected by the ``small yard, high fence" policy.

The empirical results of this paper lay a solid foundation for further analysis of the impact of China-U.S. trade friction on China’s innovation and technological progress, offering important references for China’s response to U.S. technology restrictions. Future research can proceed in the following two areas. First, systematically analyzing China-U.S. trade and technological friction requires a theoretical framework that connects international knowledge flow with trade policies. From the findings of this study, knowledge flow is not just a byproduct of international trade; it has become a key factor in the formulation of trade policies by major powers today. Second, the empirical conclusions of this paper suggest that the ``small yard, high fence" policy has persistently reduced international technology flow in certain Chinese technology fields. How will this upstream impact in the knowledge and technology domain be transmitted to downstream production processes? The influence from knowledge flow to economic impact requires some time to transmit, and future research could further discuss the economic effects of the ``small yard, high fence" policy based on this study.

\newpage


\bibliographystyle{apalike}
\bibliography{reference}



\end{document}